\documentclass{ws-procs9x6}

\begin{document}

\title{The proton structure function $F_2$ in the
resonance region}

\author{
\underbar{M.~OSIPENKO},
G.~RICCO,
M.~TAIUTI,
M.~ANGHINOLFI,
M.~BATTAGLIERI,
R.~DE~VITA,
M.~RIPANI
}

\address{Istituto Nazionale di Fisica Nucleare, Sezione di Genova, e
Dipartimento di Fisica dell'Universit\`a, 16146 Genova, Italy}

\author{
S.~SIMULA
}

\address{Istituto Nazionale di Fisica Nucleare, Sezione Roma III,
00146 Roma, Italy}

\author{
G.~FEDOTOV,
E.~GOLOVACH,
B.~ISHKHANOV,
V.~MOKEEV
}
\address{Nuclear Physics Institute of Moscow State University,
119992 Moscow, Russia}

\author{and the CLAS Collaboration}

\maketitle

\abstracts{Unique measurement of the proton structure function
$F_2$ in a wide two-dimensional region of $x$ and $Q^2$ has been
reported. The accessible kinematics covers entire resonance region
up to $W=2.5$ GeV in the $Q^2$ interval from 0.1 to 4.5 GeV$^2$.
Obtained data allowed for the first time an evaluation
of moments of the structure function $F_2$ directly from
experimental data as well as an intensive study of the Bloom-Gilman
duality phenomenon.}

\section{Introduction}\label{sec:int}
So far the inclusive electron scattering off the nucleon represents one of
the simplest and most powerful probes of the hadron structure.
Many interesting results were obtained from such experiments in
Deep Inelastic Scattering (DIS) regime, where smallness of the strong
interaction coupling constant offers a simple interpretation of
the results in terms of perturbative QCD (pQCD). These studies however,
while providing a detailed information on the partons forming the
nucleon, completely neglect the complementary and interesting phenomena
which characterize the non perturbative regime where partons are
correlated. Moreover, most of the nucleon mass is created by the
dynamical processes not seen in the DIS.

In order to study these phenomena one has to use medium or low $Q^2$ probes
providing the information on the large distance physics. In early experiments
performed in this region nearly 30 years ago at SLAC it was noted by Bloom
and Gilman\cite{BloGil} that even in presence of the peaks of the nucleon
excited states the measured structure function $F_2$ shows something
in common with the one extrapolated from DIS. In particular, the structure
function averaged over resonance bumps exhibits scaling behaviour in terms of
so called ``improved scaling variable'' $x'=x/(1+M^2x^2/Q^2)$. This
phenomenon was called duality. Moreover, the duality appears to be local,
which means that each resonance peak averaged within it's own width also
shows DIS behaviour. Later on, in the frame of pQCD,
De Rujula, Georgi and Politzer\cite{Rujula} provided a first
explanation of the Bloom-Gilman duality. Using Wilson expansion
of the product of two hadronic currents they obtained a Regge-like
decomposition of the structure function moments $M_n$:
\begin{equation}
M_n(Q^2)=\sum_{\tau=2k}^{\infty}E_{n \tau}(\mu,Q^2)
O_{n \tau}(\mu)\biggl(\frac{\mu^2}{Q^2}\biggr)^{{1 \over 2}(\tau-2)},
\label{eq:i_m1}
\end{equation}
\noindent
where $k=1,2,...$, the parameter $\mu$ is the factorization scale,
$O_{n \tau}(\mu)$ is the reduced matrix element of the local operators
with definite spin $n$ and twist $\tau$ (dimension-spin), related to
the genuine non-perturbative structure of the target and $E_{n \tau}(\mu,Q^2)$
is a dimensionless coefficient function, describing small distance behaviour,
which can be perturbatively expressed as a power expansion of the running
coupling constant $\alpha_s(Q^2)$.

The leading twist term $\tau=2$ is well established in DIS, in contrast
to higher twists, responsible for confinement and duality phenomenon. In order
to study the higher twists contribution it is essential to have a complete
set of experimental data on the structure function $F_2$ covering the entire
$x$-range at each fixed $Q^2$. But, as it was shown in Ref.\cite{Ricco2}
higher twists can be well established only with higher moments ($n>2$),
meanwhile for $M_2$ their contribution is very small even at $Q^2$
about 1 GeV$^2$. Therefore, the most interesting kinematic region is
situated at small $Q^2$ from 0 to 5 GeV$^2$ and large values of $x$,
where the higher moments dominate. The experiment described in this report
was performed at TJNAF on the CLAS detector and it covered most of this
region.

\section{Data analysis}\label{sec:dat}
\begin{figure}[ht]
\centerline{\epsfxsize=2.0in\epsfbox{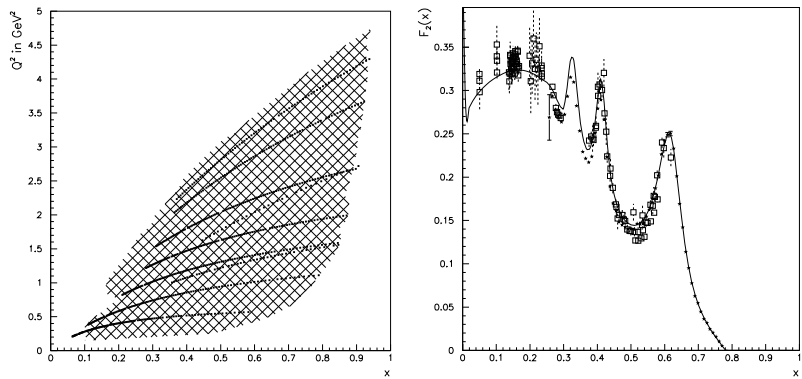}~~~\epsfxsize=2.0in\epsfbox{sf2.epsi}}
\caption{The kinematics covered by present experiment
(hatched area) in comparison with one achieved in
Ref.$^8$ on classical two arm spectrometer (left);
the obtained structure function $F_2$ at $Q^2=0.975$ GeV$^2$:
full stars - present experiment; open squares - existing
world data; solid line - parametrization from Ref.$^3$ (right). \label{fig1}}
\end{figure}

The data have been collected at TJNAF in Hall B with the
large solid angle CLAS detector~\cite{CLAS}
on a liquid hydrogen target during the electron beam running period in
February-March 1999. To cover the largest interval in $Q^2$ and $x$
data have been taken at five different electron beam energies:
$E_0=$~1.5, 2.5, 4.0, 4.2 and 4.4 $GeV$. The kinematics covered in
this experiment is shown in Fig.\ref{fig1} and compared to the one
obtained with a typical small acceptance spectrometer.

The efficiency study was based on GEANT based Monte-Carlo simulation
of the CLAS. Contamination of the background processes appeared to be
significant in some particular limits of the large kinematics covered
in the experiment. For instance, the contamination of negative pions,
high energy $e^+e^-$-pair production and fast knock-on electrons
contribution were computed and removed from the final data.
The efficiency and corrections were carefully checked comparing of the
well known elastic scattering cross section to the measured one.

The structure function $F_2$ was extracted from the inelastic
cross section using the parameterization of the function
$R(x,Q^2)\equiv\sigma_L/\sigma_T$ developed in Ref.\cite{Ricco1}
based on low $Q^2$ data from Ref.\cite{Ratio}.
%The function $R(x,Q^2)$ measured does not consider recently
This parameterization does not include the recently
measured data in the resonance region reported in these
proceedings~\cite{hc_ratio}. However the structure function
$F_2$ is not very sensitive to the value of $R$. In fact even
a 100\% systematic uncertainty on $R$ gives only few percent
uncertainty on $F_2$ in this region.

\section{Obtained results}\label{sec:res}
In order to have a precise evaluation of the structure function moments
we used a very large set of data. This set consists of recent data from CLAS
and all previous world data on the inclusive cross section of the
charged lepton-proton scattering\cite{f2-hc,f2world,csworld}. It allowed for the
first time to evaluate moments by the numerical integration of the
experimental data with almost complete coverage of entire $x$
interval by a dense set of points (see Fig.\ref{fig1} left panel).
The statistical errors of the moments are quite small, while systematic
uncertainties vary from 1 to 10\%. We performed separation of the higher
twists contribution in the frame of phenomenological
approach\cite{Ricco2,Ricco1,SIM00}. The higher twists contribution was
obtained with high precision due to the large amount of data and it will
be published on a forthcoming paper.

\begin{figure}[ht]
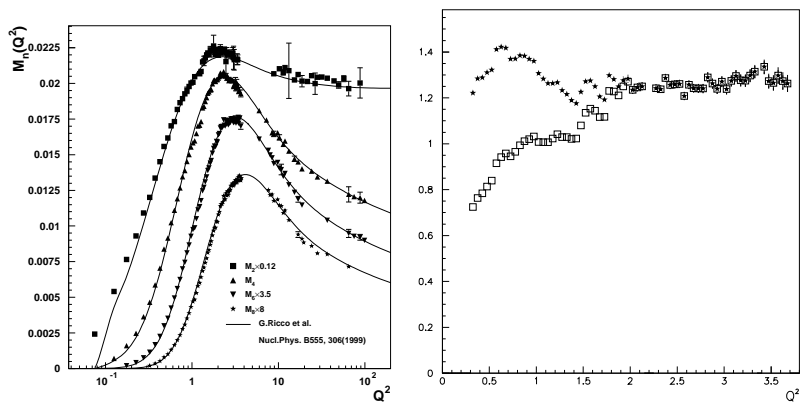

\centerline{\epsfxsize=2.0in\epsfbox{moms3.epsi}~~~\epsfxsize=2.0in\epsfbox{dual4.epsi}}
\caption{The obtained moments of the structure function $F_2$ (left);
ratio of the cross section averaged within one resonance peak width
over cross section extrapolated from DIS for the first resonance bump:
empty boxes and stars represent the ratio with and without $Q^2$ evolution
of the DIS cross section, correspondingly. \label{fig2}}
\end{figure}

To study the local duality phenomenon we performed the
comparison of measured cross sections in the resonance region with
different DIS fits\cite{GRV,MRS,CTEQ} extrapolated to lower $Q^2$
and higher $x$ values. This analysis was done according the following
procedure:
\begin{romanlist}
\item First of all for each of three clearly observed resonance
bumps ($W=1.232$, $1.52$, $1.7$ GeV) we averaged the experimental
cross section within peak width;
\item then we calculated the structure function $F_2$ from one of DIS
parameterizations\cite{GRV,MRS,CTEQ};
\item we corrected the DIS structure function according for Ref.\cite{Rujula}
on the target mass corrections;
\item the obtained structure function $F_2$ has been converted into
cross section by using the parametrization of $R=\sigma_L/\sigma_T$
already mentioned above, and the cross section has been averaged
within the same three intervals as the experimental data;
\item finally, we computed the ratio of experimental values of the averaged
cross section over ones obtained from DIS extrapolation.
\end{romanlist}
We have tried three different parameterizations\cite{GRV,MRS,CTEQ} as
well as different target mass corrections\cite{Rujula,G&P};
we found that the duality can be extended to lower momentum transfers
(about $0.5$ GeV) if no $Q^2$ evolution is applied to the structure
function below $Q^2=2$ GeV$^2$ (see Fig.\ref{fig2} right panel).

%Summarizing, it should be noted that 
In conlcusion, the phenomenon of local duality
remains still somewhat mysterious. In particular, it is unclear why
one has to average the resonance bumps formed by a set of excited states
(second peak consists of $D_{13}(1520)$ and $S_{11}(1535)$, third peak
contains $F_{15}(1680)$, $P_{13}(1720)$, $D_{13}(1700)$ and $P_{11}(1710)$)
which have different spins, masses and widths. The way one averages
the peaks is completely arbitrary, since it does not relate neither to
a particular resonance nor to a set of excited states belonging to a common
symmetry-type\cite{Isgur}. However, in the right panel of Fig.\ref{fig2}
we clearly see that duality is fulfilled at $Q^2$ above $1$ GeV$^2$.
%Suppose duality is related to the Coulomb sum rule
%(see Ref.\cite{Isgur,Sabine}) and it has to be satisfied
%rigorously. Then obtained results mean that either standard pQCD $Q^2$
%evolution of the running coupling constant $\alpha_S(Q^2)$ rather
%inappropriate at $Q^2$ below 2 GeV$^2$ or the target mass correction
%calculated according to Ref.\cite{Rujula} does not fit in the reality.
Assuming that duality is related to the Coulomb sum rule
(see Ref.\cite{Isgur,Sabine}) and that it has to be satisfied
rigorously, then the obtained results indicate that either the standard pQCD $Q^2$
evolution of the running coupling constant $\alpha_S(Q^2)$ is rather
inappropriate at $Q^2$ below 2 GeV$^2$ or that the target mass corrections
calculated according to Ref.\cite{Rujula} do not fit in the reality.

\end{document}